\documentclass[12pt]{article}

\usepackage[a4paper,left=30mm,right=20mm,top=20mm,bottom=20mm]{geometry}
\usepackage{graphics}
\usepackage{amsmath}
\usepackage{epsfig}
\usepackage{cite}

\title{Associated production of $Z$ bosons and $b$-jets at the LHC in the combined $k_T$ + collinear QCD factorization approach}
\author{S.P.~Baranov$^{1}$, H.~Jung$^{2}$, A.V.~Lipatov$^{3,4}$, M.A.~Malyshev$^{3}$}

\begin{document}

\maketitle

\begin{center}
{\it $^1$P.N.~Lebedev Physics Institute, 119991 Moscow, Russia}\\
{\it $^2$Deutsches Elektronen-Synchrotron, Notkestrasse 85, Hamburg, Germany}\\
{\it $^3$Skobeltsyn Institute of Nuclear Physics, Moscow State University, 119991 Moscow, Russia}\\
{\it $^4$Joint Institute for Nuclear Research, Dubna 141980, Moscow region, Russia}
\end{center}

\vspace{0.5cm}

\begin{center}

{\bf Abstract }

\end{center} 

\indent

We consider the production of $Z$ bosons associated with beauty 
quarks at the LHC using a combined $k_T$ + collinear QCD factorization approach,
that interpolates between small $x$ and large $x$ physics.  
Our consideration is based on the off-shell gluon-gluon fusion subprocess 
$g^* g^* \to Z Q\bar Q$ at the leading order ${\cal O}(\alpha\alpha_s^2)$
(where the $Z$ boson further decays into a lepton pair), calculated 
in the $k_T$-factorization approach, and several subleading 
${\cal O}(\alpha \alpha_s^2 )$ and ${\cal O}(\alpha \alpha_s^3 )$
subprocesses involving quark-antiquark and quark-gluon interactions,
taken into account in conventional (collinear) QCD factorization.
The contributions from double parton scattering are discussed as well.
The transverse momentum dependent (or unintegrated) 
gluon densities in a proton are derived from Catani-Ciafaloni-Fiorani-Marchesini (CCFM)
evolution equation.
We achieve reasonably good agreement with the latest data taken by CMS and 
ATLAS Collaborations. 
The comparison of our results with next-to-leading-order pQCD predictions, obtained
in the collinear QCD factorization, is presented. We discuss the uncertainties of our calculations and 
demonstrate the importance of subleading quark involving contributions in describing 
the LHC data in the whole kinematic region.

\vspace{1.0cm}

\noindent
PACS number(s): 12.38.-t, 12.38.Bx, 13.85.Qk, 14.65.-q, 14.70.Hp

\newpage
\indent

\section{Introduction} \indent

With the LHC in operation, one can access a number of "rare" processes
which could have never been systematically studied at previous accelerators.
In this article we draw attention to the associated production of $Z$ 
bosons and $b$-jets. This process involves both weak and strong
interactions and therefore serves as a complex test of Standard Model, perturbative QCD (pQCD)
and our knowledge of parton distribution functions in a proton. Similarly to the $W + c$ and $W + b$
processes considered earlier\cite{1,2} it probably provides an arena for double 
parton scattering (DPS), now widely discussed in the literature. We wish to clarify 
this point in our paper.
Besides that, this process constitutes a substantial background in studying the 
associated production of Higgs and $Z$ bosons, where the Higgs boson is 
identified via its decay into a $b\bar{b}$ pair\cite{3,4,5}. A number of physics 
scenarios beyond Standard Model also refer to final states containing $Z$ bosons and 
beauty quarks\cite{6,7,8}, thus making the related studies important and topical.

Our present study is greatly stimulated by the recent ATLAS measurements\cite{9} of the total 
and differential production cross sections of $Z$ bosons associated with beauty quark 
jets at $\sqrt{s} = 7$~TeV accompanied by the CMS measurements\cite{10} of kinematic correlations 
between $Z$ bosons and $b$-hadrons at $\sqrt{s} = 7$~TeV. 
We investigate these processes in the framework
of a combined QCD approach, based on the $k_T$-factorization
formalism\cite{11,12} in the small-$x$ domain and conventional (collinear) QCD factorization at large Bjorken $x$.
Doing so, we employ the $k_T$-factorization approach 
to calculate the leading contributions from the off-shell
gluon-gluon fusion $g^* g^* \to Z Q\bar Q$ and,
to extend the consideration to the whole kinematic range,
take into account several subleading quark-involved
subprocesses using collinear QCD factorization.
The $k_T$-factorization approach has certain technical advantages in the 
ease of including higher-order radiative corrections that can be taken into account 
in the form of transverse momentum dependent (TMD) 
parton distributions\footnote{See reviews\cite{13} for more 
information.}. This approach has become a widely 
exploited tool and it is of interest and importance to test it in as many cases as 
possible. Closely related to this is the selection of TMD parton densities best 
suited to describe the data. These tasks form the major goal of our article. 

The outline of the paper is the following. In Section~2 we describe our approach 
and parameter setting. In Section~3 we present the results of our calculations and 
confront them with the available data. Our conclusions are summarised in Section~4.

\section{The model} \indent

Let us start from a short review of calculation steps.
The leading contribution comes from the ${\cal O}(\alpha \alpha_s^2)$ off-shell gluon-gluon fusion subprocess:
\begin{equation} \label{gg-ZQQ}
  g^*(k_1) + g^*(k_2) \to Z(p) + b(p_1) + \bar b(p_2)
\end{equation}

\noindent
where the four-momenta of all particles are given in
the parentheses.
The corresponding gauge-invariant off-shell amplitude was calculated earlier\cite{14,15}
and implemented into the Monte-Carlo event generator \textsc{cascade}\cite{16}.
All the details of these calculations have been explained\cite{14,15}, we only 
mention here that the standard QCD Feynman rules were employed with the only exception 
that the initial off-shell gluon spin density matrix was determined according to 
the $k_T$-factorization prescription\cite{11,12}:
\begin{equation}
  \sum \epsilon^\mu (k_i) \epsilon^{* \nu} (k_i) = { {\mathbf k}_{iT}^\mu {\mathbf k}_{iT}^\nu \over {\mathbf k}_{iT}^2}
\end{equation}

\noindent 
with ${\mathbf k}_{i\,T}$ being the component of the gluon momentum $k_i$ (with $i = 1$ or $2$) perpendicular
to the beam axis ($k_i^2 = - {\mathbf k}_{iT}^2 \neq 0$).
In the collinear limit ${\mathbf k}_{iT}^2 \to 0$ this expression converges to the ordinary
one after averaging on the azimuthal angle.

In order to fully reproduce the experimental setup\cite{9,10}, 
we simulate the subsequent decay $Z\to l^+l^-$ incorporated with the production step at the amplitude level.
Then, the $Z$ boson propagator is parametrised in Breit-Wigner form %$1/(p_Z^2-m_Z^2-i\Gamma m_Z)$
with mass $m_Z = 91.1876$~GeV and total decay width $\Gamma_Z = 2.4952$~GeV\cite{17}.
The role of virtual photons in the $Z$ boson resonance region is
found to be small: it makes not more than a $2$\% or $3$\% correction (including the 
$Z/\gamma^*$ interference effects). This is much less than the scale uncertainty 
of the main subprocess (see Section~3), and, therefore, is neglected in our analysis.

In addition to off-shell gluon-gluon fusion, we take into account several subprocesses involving 
quarks in the initial state. These are the flavor excitation at ${\cal O}(\alpha \alpha_s^2)$:
\begin{equation} \label{qQ-ZqQ}
  q(k_1) + b(k_2) \to Z(p) + q(p_1) + b(p_2);
\end{equation}

\noindent
the quark-antiquark annihilation at ${\cal O}(\alpha \alpha_s^2)$:
\begin{equation} \label{qq-ZQQ}
  q(k_1) + \bar q(k_2) \to Z(p) + b(p_1) + \bar b(p_2);
\end{equation}

\noindent
and the quark-gluon scattering at ${\cal O}(\alpha \alpha_s^3)$:
\begin{equation} \label{qg-ZqQQ}
  q(k_1) + g(k_2) \to Z(p) + q(p_1) + b(p_2) + \bar b(p_3).
\end{equation}

\noindent
Quark densities are typically much lower than the gluon density at the LHC conditions;
however, these processes may become important at very large transverse momenta
(or, respectively, at large parton longitudinal momentum fraction $x$, which is needed to
produce large $p_T$ events) where the quarks are less suppressed or can even dominate 
over the gluon density. Here we find it reasonable to rely upon collinear 
Dokshitzer-Gribov-Lipatov-Altarelli-Parisi (DGLAP) factorization scheme\cite{18}, 
which provides better theoretical grounds in the large-$x$ region.
So, we consider a combination of two techniques with each of them being used at the 
kinematic conditions where it is best suitable (gluon-induced subprocess~(1)
at small $x$ and quark-induced subprocesses (3) --- (5) at large $x$ values).
For the flavor excitation and the quark-antiquark annihilation
we apply the on-shell limit of formulas obtained 
earlier\cite{19} supplementing them by the $Z$ boson decays.
The amplitude of quark-gluon scattering subprocess
can be easily derived from the gluon-gluon fusion one.

As usual, to calculate the contributions of quark-induced subprocesses~(3) --- (5) one has to 
convolute the corresponding partonic cross sections $d\hat \sigma_{ab}$ with the 
conventional parton distribution functions $f_a(x,\mu^2)$ in a proton:
\begin{equation}
  \sigma = \int dx_1 dx_2 \, d\hat \sigma_{ab}(x_1,x_2,\mu^2) f_a(x_1,\mu^2)  f_b(x_2,\mu^2),
\end{equation}

\noindent
where indices $a$ and $b$ denote quark and/or gluon, 
$x_1$ and $x_2$ are the fractions of longitudinal momenta of colliding protons
and $\mu^2$ is the hard scale.
In the case of off-shell gluon-gluon 
fusion we employ the $k_T$-factorization formula:
\begin{equation}
  \sigma = \int dx_1 dx_2 \, d{\mathbf {k}_{1T}^2} d{\mathbf {k}_{2T}^2} \, d\hat \sigma_{gg}^*(x_1,x_2,{\mathbf {k}_{1T}^2},{\mathbf {k}_{2T}^2},\mu^2) f_g(x_1,{\mathbf {k}_{1T}^2},\mu^2)  f_g(x_2,{\mathbf {k}_{2T}^2},\mu^2),
\end{equation}

\noindent
where $f_g(x,{\mathbf {k}_{T}^2},\mu^2)$ is the TMD gluon density in a proton.
To obtain the latter we use a numerical solution of the CCFM
equation\cite{20}. It provides a suitable tool as it smoothly interpolates 
between the small-$x$ Balitsky-Fadin-Kuraev-Lipatov (BFKL)\cite{21} gluon dynamics and large-$x$ DGLAP one. We adopt 
the latest JH'2013 parametrization\cite{22}, taking JH'2013 set~2 as the default choice.
The corresponding TMD gluon density has been fitted to high-precision DIS data on the 
proton structure functions $F_2(x,Q^2)$ and $F_2^c(x,Q^2)$. The fit was based on TMD matrix elements and 
involves two-loop strong coupling constant, kinematic consistency constraint\cite{23,24} 
and non-singular terms in the CCFM gluon splitting function\cite{25}.
For the conventional quark and gluon densities we use the MSTW'2008 (LO) set\cite{26}.

Throughout this paper, all calculations are based on the following parameter setting.
In the collinear QCD factorization case we use one-loop 
running strong and electroweak coupling constants with $n_f = 4$ massless quark flavors 
and $\Lambda_{\rm QCD} = 200$~MeV; the factorization and renormalization scales are 
both set equal to the $Z$ boson transverse mass, so that 
we have $\alpha_s(m^2_Z) = 0.1232$ and $\alpha(m^2_Z) = 1/128$.
In the $k_T$-factorization case we use a two-loop expression for 
the strong coupling constant (as it was originally done in the fit\cite{22}) and
define the factorization scale as $\mu^2_F = \hat{s}  + {\mathbf Q}_T^2$ 
with $\hat s$ and ${\mathbf Q}_T^2$ being the 
subprocess invariant energy and the net transverse momentum of the initial off-shell 
gluon pair, respectively. The latter definition of $\mu_F$ is unusual and is dictated
by the CCFM evolution algorithm\cite{22}.
The $b$-quark mass and Weinberg mixing angle were set to 
$m_b = 4.75$~GeV and $\sin^2\theta_W = 0.2312$\cite{17}.
When necessary, $b$-quarks were converted into $b$-hadrons using Peterson 
fragmentation function\cite{27} with $\epsilon_b = 0.006$.

We close our consideration with DPS contributions where we apply a simple factorization
formula (for details see the reviews\cite{28,29,30} %\cite{43,44}
and references therein):
\begin{equation}
  \sigma_{\rm DPS} (Z + b + \bar b) = {\sigma(Z) \, \sigma (b + \bar b) \over \sigma_{\rm eff}}, 
\end{equation}

\noindent
where $\sigma_{\rm eff}$ is a normalization constant which incorporates all 
"DPS unknowns" into a single phenomenological parameter. A numerical value of 
$\sigma_{\rm eff} \simeq 15$~mb was earlier obtained from fits to $pp$ and $p\bar{p}$ 
data. %\cite{45,46,47,48,49}. 
This will be taken as the default value throughout the paper.
The calculation of inclusive cross sections $\sigma(b + \bar{b})$ 
and $\sigma(Z)$ is straightforward and needs no special
explanations. Here we strictly follow the approach described earlier\cite{31,32,33}.

The multidimensional phase space integration was performed by means of the
Monte Carlo technique, using the routine \textsc{vegas}\cite{34}.
In the next section we confront our predictions with the latest LHC data.

\section{Numerical results} \indent

This section presents a detailed comparison between theoretical calculations
and recent LHC data. The essential measurements have been carried out by the
ATLAS\cite{9} and CMS\cite{10} Collaborations and refer to the following
categories: $Z$ bosons produced in association with one beauty jet,
$Z$ bosons produced in association with two beauty jets and $Z$ bosons produced 
in association with explicitly reconstructed $b$-hadrons.
In addition to the above, the ATLAS Collaboration has presented\cite{9} inclusive cross sections 
for $Z$ bosons associated with any number of $b$-jets. We do not analyse events 
of this kind in the present study and only concentrate on the production of $Z$ 
bosons with one or two $b$-jets.

\subsection{Production of $Z$ bosons in association with one $b$-jet} \indent

The ATLAS Collaboration has collected the data\cite{9} at $\sqrt s = 7$~TeV. Both 
leptons originating from the $Z$ boson decay are required to have $p_T^l > 20$~GeV 
and $|\eta^l| < 2.4$, the lepton pair invariant mass lies in the interval $76 < M^{ll} < 106$~GeV,
the beauty jets are required to have $p_T^b > 20$~GeV and $|\eta^b| < 2.4$.

We confront our predictions with the available data in Figs.~1 and~2.
To estimate the theoretical uncertainties in the quark-involving subprocesses~(3) --- (5),
calculated using the collinear QCD factorization, 
we have varied the scales $\mu_R$ and $\mu_F$ by a factor of $2$ around their default values.
In the $k_T$-factorization approach, employed for off-shell gluon-gluon fusion subprocess~(1), 
the scale uncertainties were estimated by using the gluon 
densities JH'2013 set 2$+$ and JH'2013 set 2$-$ instead of default density JH'2013 set 2. 
These two sets refer to the varied hard
scales in the strong coupling constant $\alpha_s$ in the off-shell amplitude: 
JH'2013 set 2$+$ stands for $2\mu_R$, while JH'2013  set 2$-$ refers to $\mu_R/2$ (see\cite{22} for 
more information). 
The estimated scale uncertainties are shown as shaded bands. 
As one can see, we achieve reasonably good agreement with the ATLAS data\cite{9} within 
the experimental and theoretical uncertainties, although we observe some  
underestimation of these data at high $p_T^Z$ and a slight 
overestimation at small transverse momenta.
The slight overestimation of the data at low $p_T^Z$ can probably 
be attributed to the TMD gluon density used, since the region $p_T^Z < 100$~GeV is fully 
dominated by off-shell gluon-gluon fusion, as it is demonstrated in Fig.~2.
The rapidity distribution is well described practically everywhere. 
The NLO pQCD calculations\footnote{We take them from ATLAS publication\cite{9}.}, 
performed using \textsc{mcfm} routine\cite{35}, tend to slightly overestimate our 
predictions and better decribe the data at large transverse momenta.

To investigate the importance of $k_T$-factorization, we have repeated
the calculation using collinear QCD factorization for all considered
subprocesses (dash-dotted histograms in Fig.~1). We find that these effects are significant 
at low and moderate $p_T^Z$ (up to $p_T^Z \sim 100$~GeV), where the off-shell gluon-gluon fusion 
dominates. The effect of using $k_T$-factorization for gluon-dominated processes is clearly demonstrated in Fig.~2.
The quark-initiated subprocesses~(3) --- (5) become important only 
at high transverse momenta, where the typical $x$ values are large, and
that supports using of the DGLAP quark and gluon dynamics for these subprocesses (see Fig.~2).
The subprocesses~(3) --- (5) are important to achieve an adequate description of the 
data in the whole $p_T^Z$ region.

The estimated DPS contributions are found to be small in the considered kinematic 
region.
Some reasonable variations in $\sigma_{\rm eff} \simeq 15 \pm 5$~mb
would affect DPS predictions, though without changing our basic conclusion.
We note also that scale uncertainties of the CCFM-based predictions are comparable 
with the ones of NLO pQCD calculations.

\subsection{Production of $Z$ bosons in association with two $b$-jets} \indent

The data provided by the ATLAS\cite{9} Collaboration refer to 
the same energies and kinematic restrictions as in the previous subsection. The 
observables shown by the ATLAS Collaboration are the $Z$ boson transverse momentum $p_T^Z$ 
and rapidity $y^Z$, invariant mass of the $b$-jet pair $M^{bb}$ 
and angular separation in $\eta - \phi$ plane between the jets $\Delta R^{bb}$.
The latter is useful to identify the 
contributions where scattering amplitudes are dominated by terms 
involving gluon splitting $g \to Q + \bar Q$.

The results of our calculations are shown in Fig.~3 in comparison with the 
ATLAS data\cite{9}. As one can see, our results describe the data reasonably well 
within the experimental and theoretical uncertainties, 
although some tendency to slightly underestimate the data 
at high transverse momentum $p_T^Z$ and large $M^{bb}$ can be seen.
The role of off-shell gluon-gluon fusion subprocess
is a bit enhanced here compared to the case of $Z + b$ production
because the quark-antiquark annihilation subprocess~(4) gives a 
negligible contribution and gluon splitting subprocess~(5)
populates mainly at low $\eta - \phi$ distances $\Delta R^{bb}$.
This subprocess is complementary to the one\cite{36} where quark-gluon scattering $q + g^*$ was dominant.
The estimated DPS contribution is small and can play a role at low $p_T^Z$ only.
The NLO pQCD calculations, performed 
using \textsc{mcfm} program\footnote{We take them from ATLAS publication\cite{9}.}, tend to slightly underestimate the 
ATLAS data at low $\Delta R^{bb}$ and $M^{bb}$, although provide 
better description of the data at large transverse 
momentum $p_T^Z$ and invariant mass $M^{bb}$.

\subsection{Production of $Z$ bosons in association with two $b$-hadrons} \indent

In the measurements reported by CMS Collaboration\cite{10}, both 
$b$-hadrons have been identified explicitly by their full decay reconstruction.
This data sample allows to study the production properties of a $Zb\bar{b}$
system even in the region of small angular seperation between the $b$ quarks
(where the usual jet analysis is not possible as the jets would overlap). 
In a specific subsample, an additional cut on the $Z$ boson transverse momentum
is applied, $p_T^Z > 50$~GeV.
The CMS Collaboration described the angular configuration of the $Zb\bar{b}$ system
in terms of spatial (in $\eta - \phi$ plane) and azimuthal separation between the $b$-hadrons
$\Delta R^{bb}$ and $\Delta\phi^{bb}$, spatial separation $\min\Delta R^{Zb}$ between
the $Z$ boson and closest $b$-hadron and the asymmetry in the $Zb\bar{b}$ system defined as
\begin{equation}
  A^{Zbb}={{\max \Delta R^{Zb}-\min \Delta R^{Zb}} \over{\max\Delta R^{Zb} +\min\Delta R^{Zb}}},
\end{equation}

\noindent
where $\max\Delta R^{Zb}$ is the distance between the $Z$ boson and remote $b$-hadron.
The correlation observables are useful to identify the different production 
mechanisms (or specific higher-order corrections). For example, 
low $\min\Delta R^{Zb}$ identifies $Z$ bosons in the vicinity 
of one of the $b$-hadron ($Z$ bosons promptly radiated from $b$-quarks), 
small $\Delta\phi_{bb}$ indicates gluon to quark splitting $g\to Q{+}\bar{Q}$.
Moreover, while the configurations where the two $b$-hadrons are 
emitted symmetrically with respect to the $Z$ directions
leads to a zero value of $A^{Zbb}$ assymetry, the additional final-state gluon 
radiation results in a non-zero one, that provides us with the possibility to test the 
high-order pQCD corrections.

Our predictions are shown in Figs.~4 and~5 in comparison with the CMS data\cite{10}.
As one can see, our results with default $b$-quark fragmentation parameters 
reasonably well describe the data
within the theoretical and experimental uncertainties.
To estimate an additional uncertainty
coming from the $b$-quark fragmentation, we repeated our calculations with
varied shape parameter $\epsilon_b = 0.003$ (not shown), which 
is often used in NLO pQCD calculations.
We find that the predicted cross sections (in the considered $p_T$ region) 
are larger for smaller $\epsilon_b$ values.
However, the typical dependence of numerical predictions on the 
fragmentation scheme is much smaller than the scale uncertainties 
of our calculations.
The NLO pQCD predictions, obtained using the a\textsc{mc@nlo}\cite{37} event 
generator\footnote{We take them from CMS publication\cite{10}.},
are rather close to our results.

\section{Conclusions} \indent

We have considered the associated $Z$ boson and beauty quark production at the LHC conditions. 
The calculations were done in a "combined"
scheme employing both the $k_T$-factorization and collinear factorization in QCD, 
with each of them used in the kinematic conditions of its best reliability.
The dominant contribution is represented by the gluon-gluon fusion subprocess
$g^*g^*\to Z b\bar{b}$ with $Z$ boson further decaying into a lepton pair.
This subprocess is entirely (for the first time) calculated in the $k_T$-factorization
approach.
A number of subleading subprocesses contributing at ${\cal O}(\alpha \alpha_s^2)$ 
and ${\cal O}(\alpha \alpha_s^3)$ have been considered in the conventional collinear scheme.

Using the TMD gluon densities derived from the CCFM evolution 
equation, we have achieved reasonably good agreement between our theoretical 
predictions and latest CMS and ATLAS experimental data collected at $\sqrt s = 7$. 
We find that the (formally subleading) quark-involving subprocesses become
especially important at high transverse momenta and are necessary to describe 
the data in the whole kinematic range.
Our estimations of the double parton scattering show that the latter is unimportant.
This conclusion is also confirmed by the fact that our single parton scattering  
calculations show no room for additional contributions when compared to the 
ATLAS and CMS data.

\section{Acknowledgements} \indent

We thank F.~Hautmann, G.I.~Lykasov and S.~Turchikhin for very useful discussions
and remarks. 
This research was supported in part by RFBR grant 16-32-00176-mol-a and
grant of the President of Russian Federation NS-7989.2016.2.
We are grateful to DESY Directorate for the
support in the framework of Moscow --- DESY project on Monte-Carlo implementation for
HERA --- LHC. M.A.M. was also supported by a grant of the foundation for
the advancement of theoretical physics "Basis" 17-14-455-1.

\newpage 

\begin{figure}
\begin{center}
\includegraphics[width=7cm]{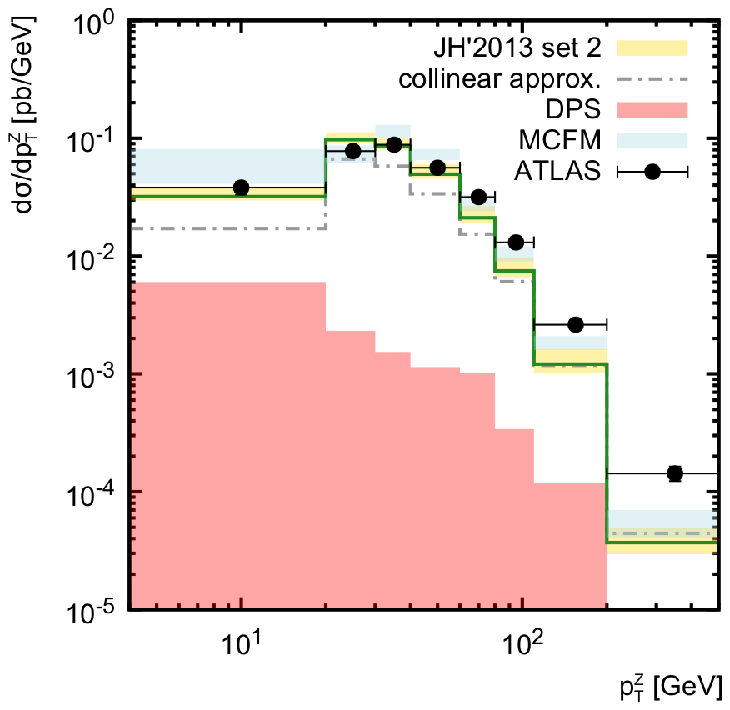}
\hspace{1cm}
\includegraphics[width=6.9cm]{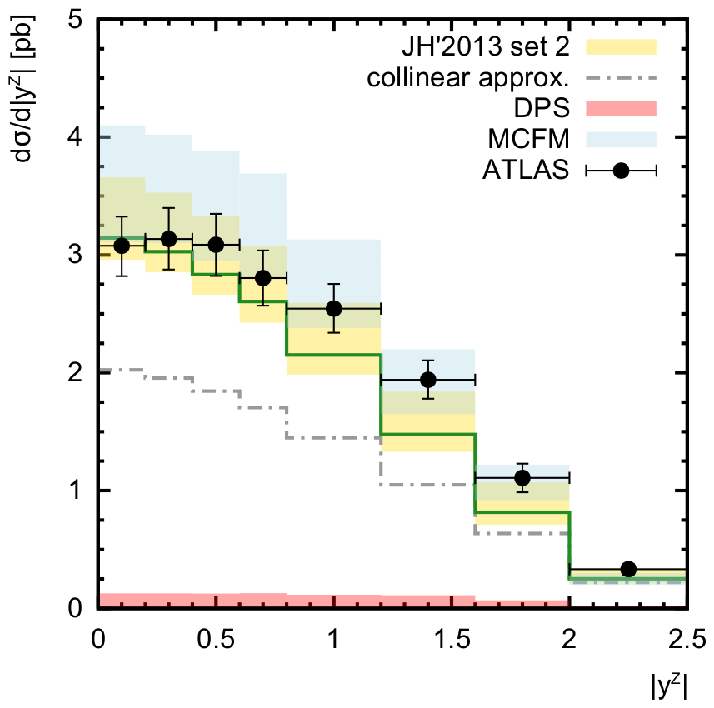}
\caption{Associated $Z + b$ production cross section 
at $\sqrt s = 7$~TeV presented as a function of the $Z$ boson transverse momentum (left panel)
or rapidity (right panel). The solid histograms show our predictions at the default scale while
shaded bands correspond to scale variations described in the text.
The dash-dotted histograms correspond to the collinear limit of our calculations. 
The estimated DPS contributions and \textsc{mcfm}\cite{35} predictions (taken from\cite{9})
are shown additionally. The data are from ATLAS\cite{9}.}
\label{fig1}
\end{center}
\end{figure}

\begin{figure}
\begin{center}
\includegraphics[width=7cm]{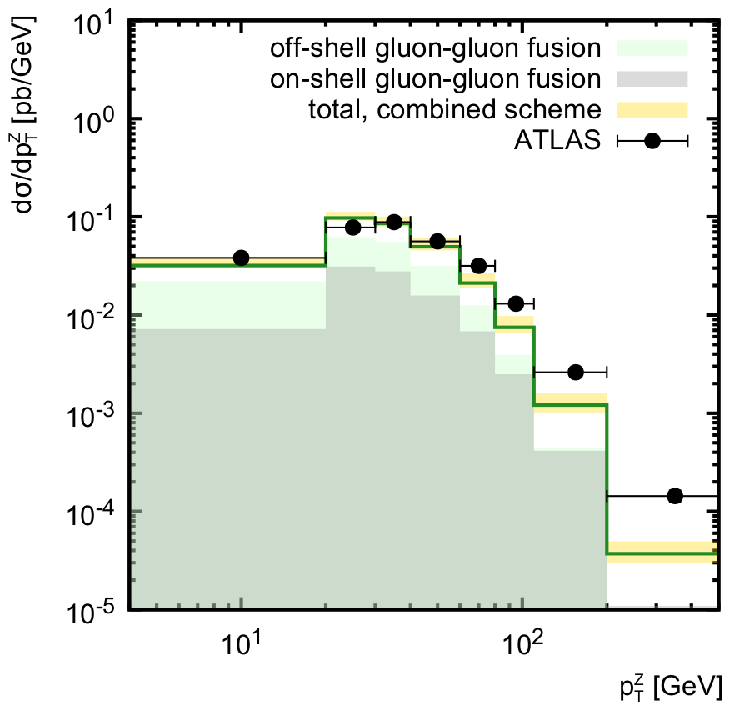}
\hspace{1cm}
\includegraphics[width=6.9cm]{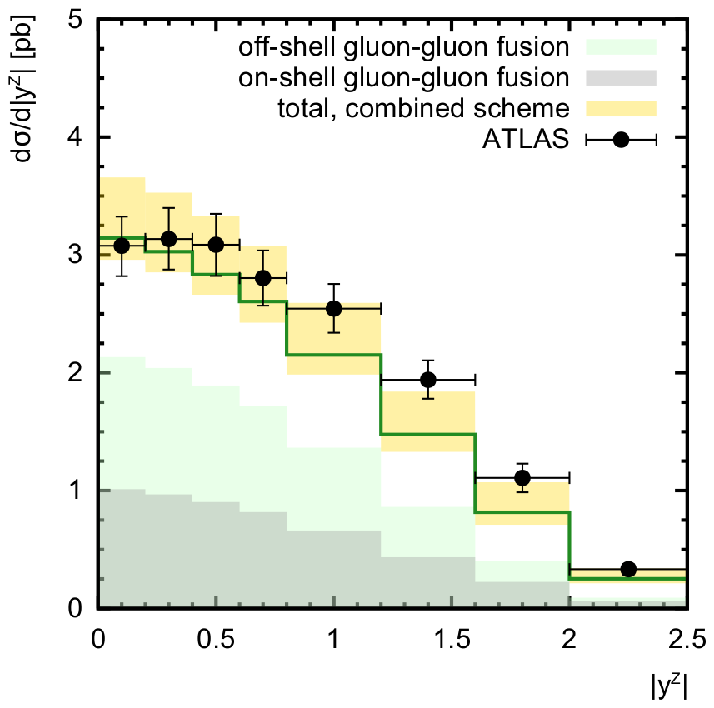}
\caption{The off-shell gluon-gluon fusion contribution to the 
associated $Z + b$ production at $\sqrt s = 7$~TeV. 
The on-shell limit of our calculations is shown additionally.
The data are from ATLAS\cite{9}.}
\label{fig2}
\end{center}
\end{figure}

\begin{figure}
\begin{center}
\includegraphics[width=7cm]{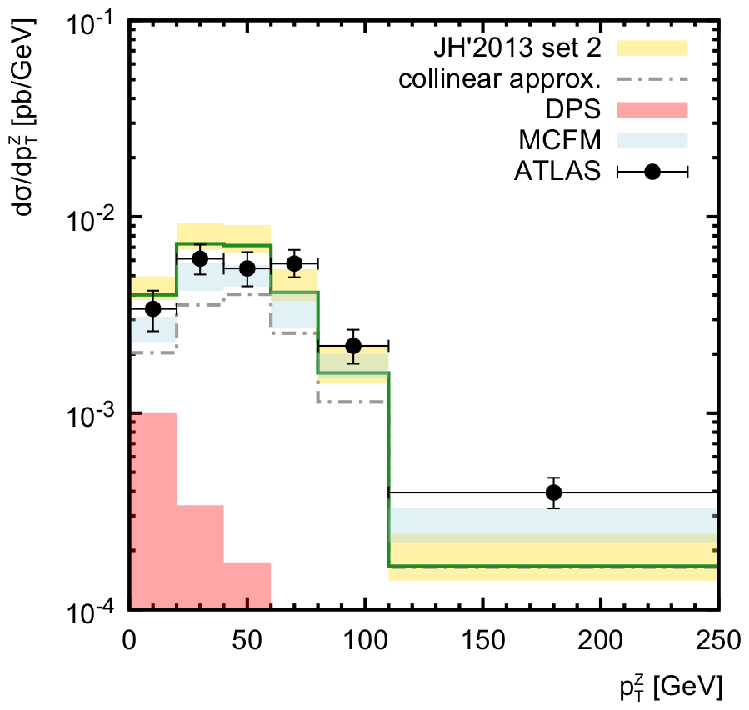}
\hspace{1cm}
\includegraphics[width=6.9cm]{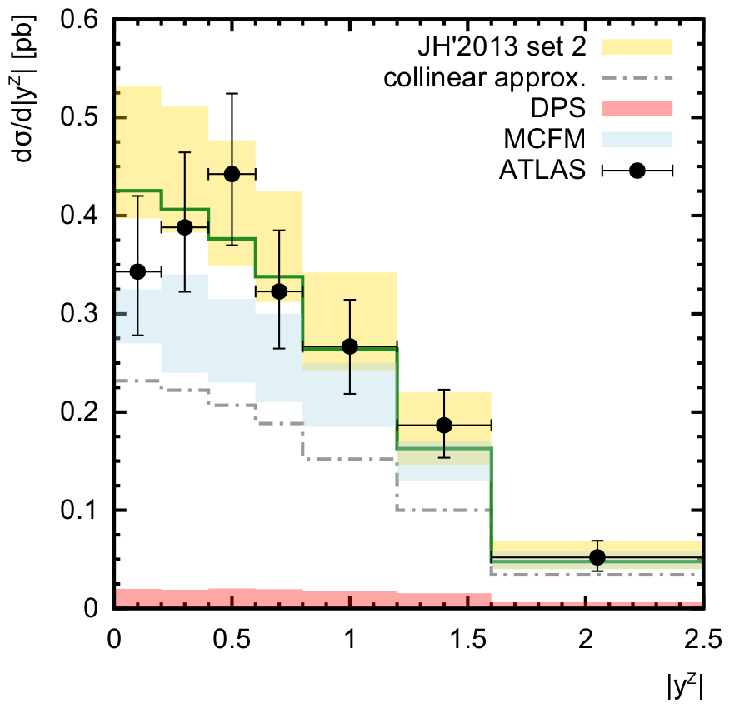}
\includegraphics[width=7cm]{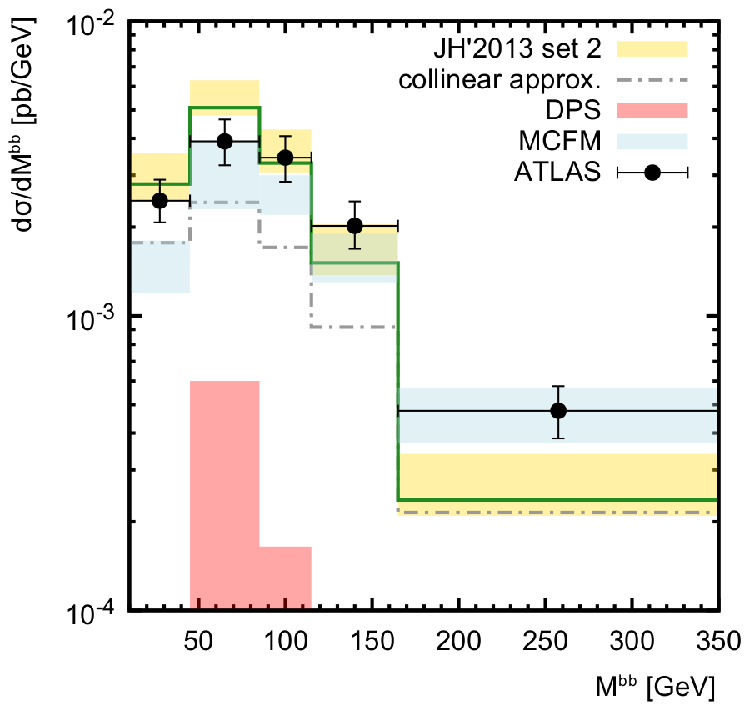}
\hspace{1cm}
\includegraphics[width=6.9cm]{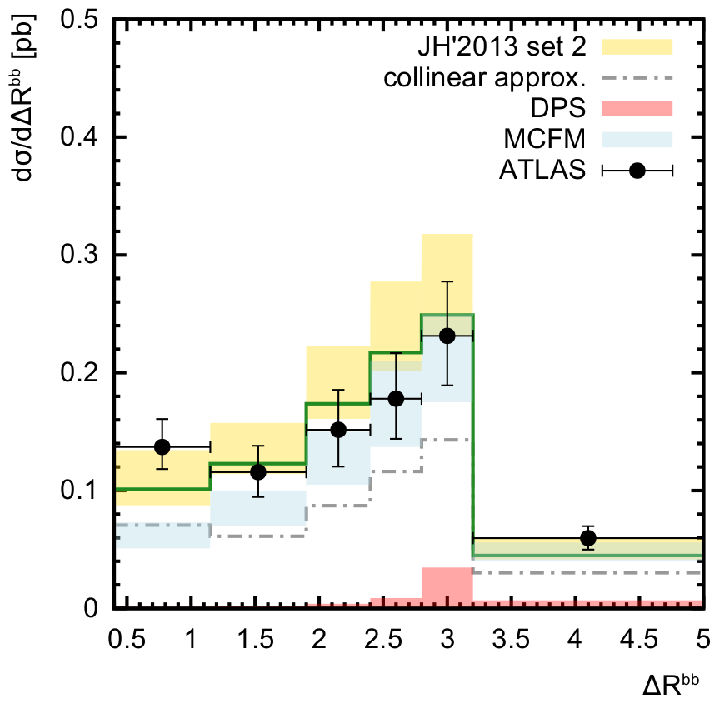}
\caption{Associated production of a $Z$ boson with 
two beauty jets at $\sqrt s = 7$~TeV calculated as a 
function of the $Z$ boson transverse momentum, rapidity, invariant mass 
of the $b$-jet pair and angular separation between the jets.
Notation of the histograms is the same as in Fig.~1.
The data are from ATLAS\cite{9}. The \textsc{mcfm}\cite{35} predictions are taken from\cite{9}.}
\label{fig3}
\end{center}
\end{figure}

\begin{figure}
\begin{center}
\includegraphics[width=7cm]{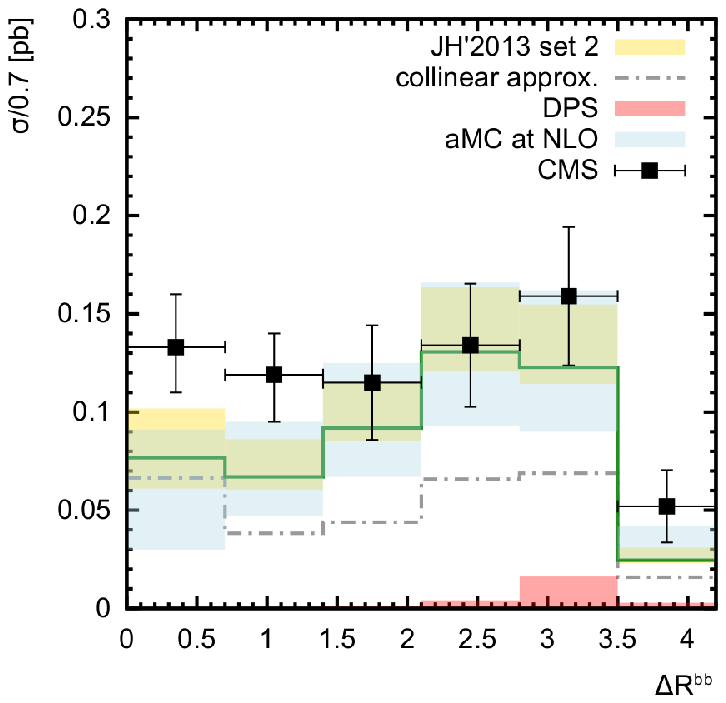}
\hspace{1cm}
\includegraphics[width=7cm]{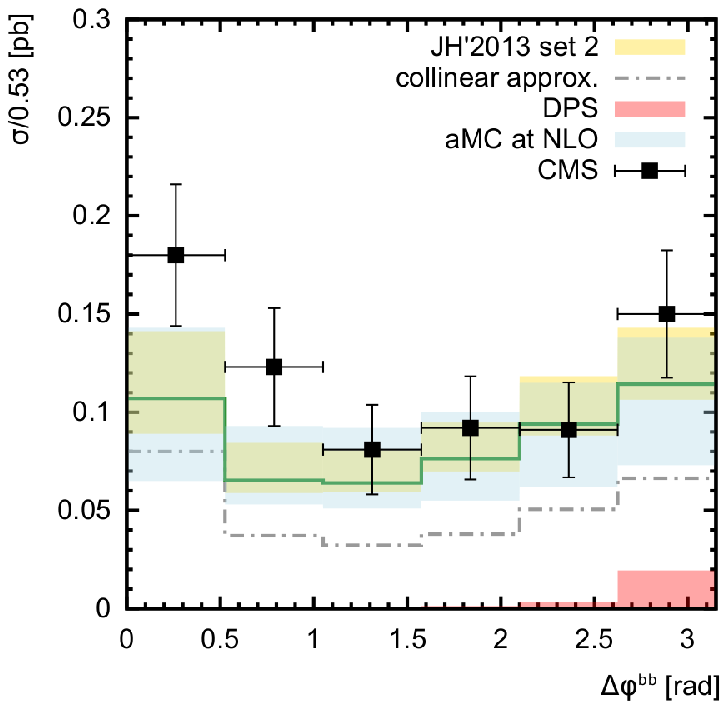}
\includegraphics[width=6.9cm]{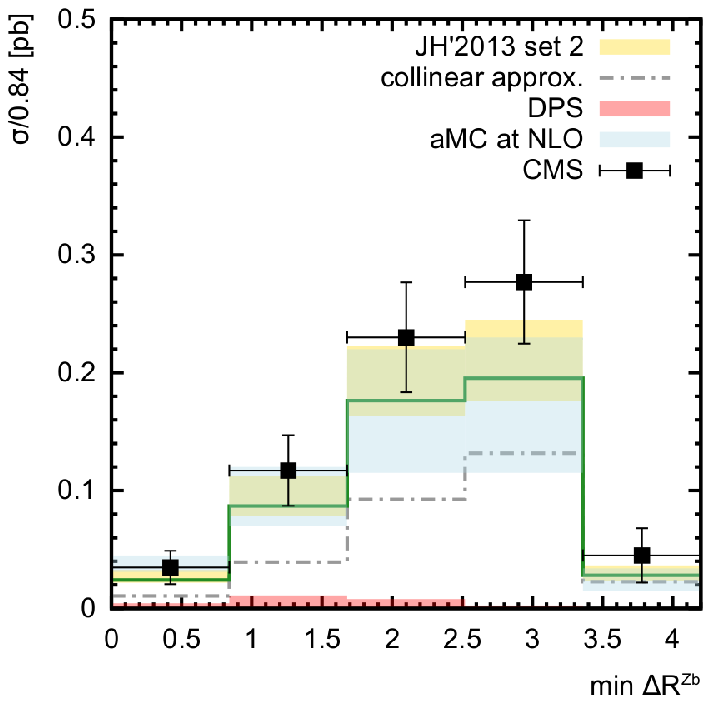}
\hspace{1cm}
\includegraphics[width=7cm]{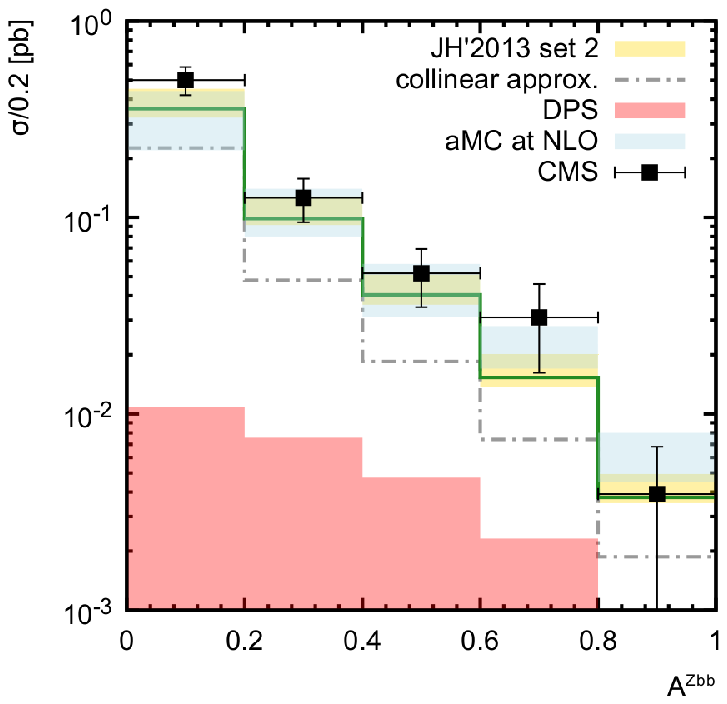}
\caption{Associated production of a $Z$ boson and two
$b$-hadrons at $\sqrt s = 7$~TeV.
Notation of the histograms is the same as in Fig.~1.
The data are from CMS\cite{10}. The a\textsc{mc@nlo}\cite{37} predictions are taken from\cite{10}.}
\label{fig4}
\end{center}
\end{figure}

\begin{figure}
\begin{center}
\includegraphics[width=7cm]{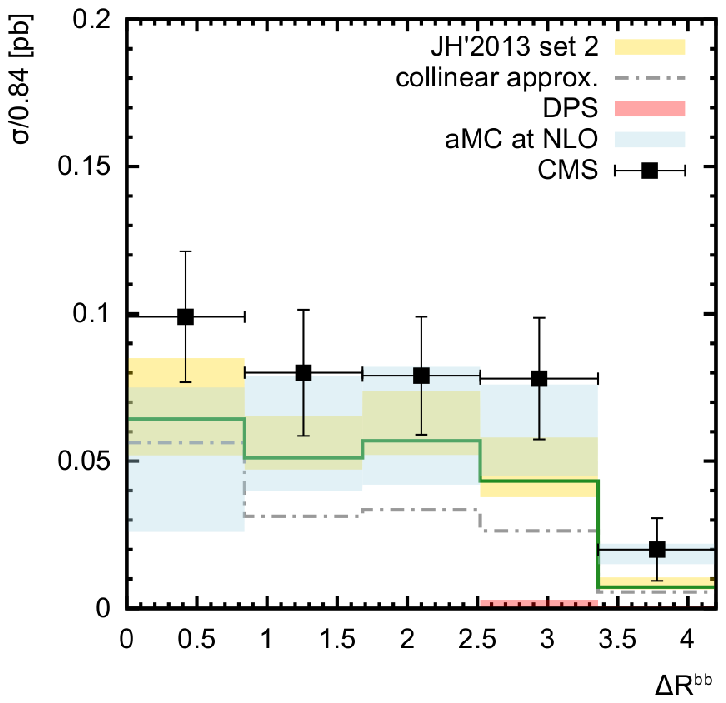}
\hspace{1cm}
\includegraphics[width=7cm]{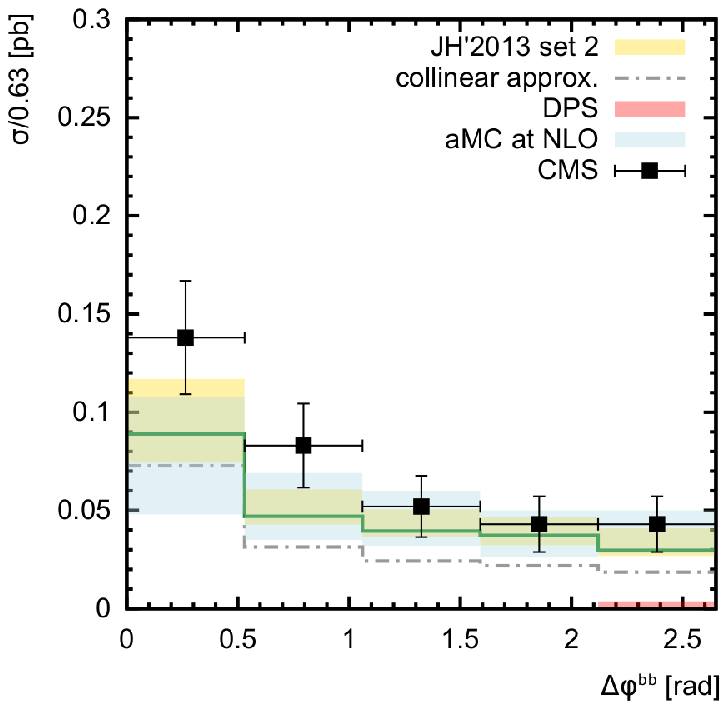}
\includegraphics[width=6.9cm]{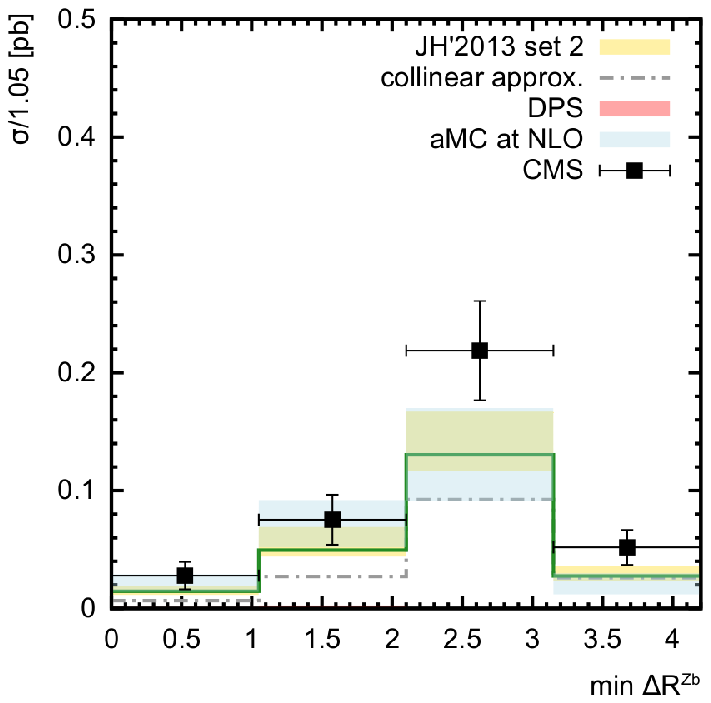}
\hspace{1cm}
\includegraphics[width=7cm]{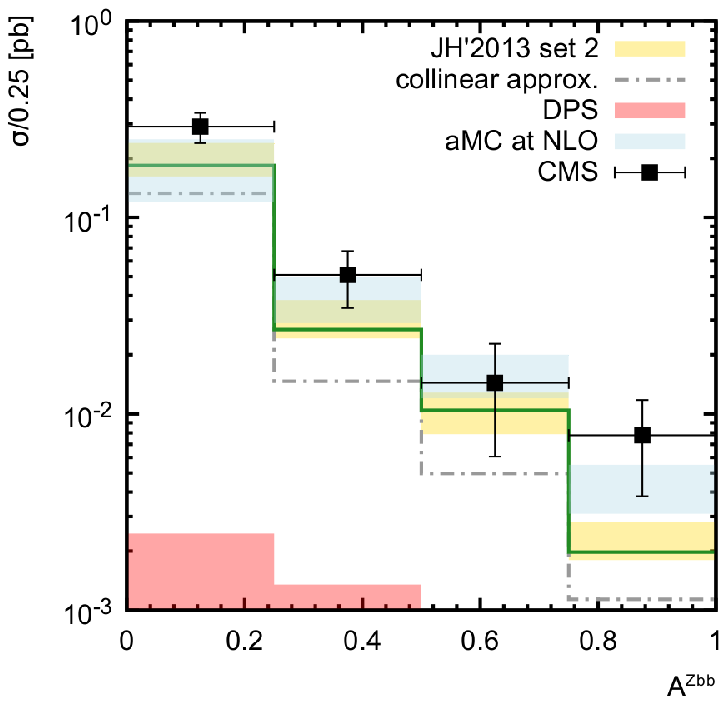}
\caption{Associated production of a $Z$ boson and two
$b$-hadrons at $\sqrt s = 7$~TeV under 
additional kinematical cut on the $Z$ boson transverse momentum $p_T^Z > 50$~GeV.
Notation of the histograms is the same as in Fig.~1.
The data are from CMS\cite{10}. The a\textsc{mc@nlo}\cite{37} predictions are taken from\cite{10}.}
\label{fig5}
\end{center}
\end{figure}


\begin{thebibliography}{37}

\bibitem{1} S.P.~Baranov, A.V.~Lipatov, M.A.~Malyshev, A.M.~Snigirev, N.P.~Zotov, Phys. Lett. B {\bf 746}, 100 (2015).
\bibitem{2} S.P.~Baranov, A.V.~Lipatov, M.A.~Malyshev, A.M.~Snigirev, N.P.~Zotov, Phys. Rev. D {\bf 93}, 094013 (2016).         
\bibitem{3} J.-M.~Gerard, M.~Herquet, Phys. Rev. Lett. {\bf 98}, 251802 (2007).
\bibitem{4} S.-de.~Visscher, J.-M.~Gerard, M.~Herquet, V.~Lemaintre, F.~Maltoni, JHEP {\bf 08}, 042 (2009).
\bibitem{5} R.-Dermisek, J.F.~Gunion, Phys. Rev. D {\bf 79}, 055014 (2009).
\bibitem{6} B.~Holdom, W.S.~Hou, T.~Hurth, M.L.~Mangano, S.~Sultansoy, G.~Unel, PMC Phys. A {\bf 3}, 4 (2009).
\bibitem{7} L.J.~Hall, D.~Pinner, J.T.~Ruderman, JHEP {\bf 04}, 131 (2012).
\bibitem{8} D.~Choudhury, T.M.P.~Tait, C.E.M.~Wagner, Phys. Rev. D {\bf 65}, 053002 (2002).
\bibitem{9} ATLAS Collaboration, JHEP {\bf 10}, 141 (2014).% Z + b/bb 7 TeV
\bibitem{10} CMS Collaboration, JHEP {\bf 12}, 039 (2013). % Z + BB 7 TeV
\bibitem{11} L.V.~Gribov, E.M.~Levin, M.G.~Ryskin, Phys. Rep. {\bf 100}, 1 (1983);\\
  E.M.~Levin, M.G.~Ryskin, Yu.M.~Shabelsky, A.G.~Shuvaev, Sov. J. Nucl. Phys. {\bf 53}, 657 (1991).
\bibitem{12} S.~Catani, M.~Ciafaloni, F.~Hautmann, Nucl. Phys. B {\bf 366}, 135 (1991);\\
  J.C.~Collins, R.K.~Ellis, Nucl. Phys. B {\bf 360}, 3 (1991).
\bibitem{13} B.~Andersson {\sl et al.} (Small-$x$ Collaboration), Eur. Phys. J. C {\bf 25}, 77 (2002);\\
  J.~Andersen {\sl et al.} (Small-$x$ Collaboration), Eur. Phys. J. C {\bf 35}, 67 (2004);\\
  J.~Andersen {\sl et al.} (Small-$x$ Collaboration), Eur. Phys. J. C {\bf 48}, 53 (2006);\\
  R.~Angeles-Martinez, A.~Bacchetta, I.I.~Balitsky, D.~Boer, M.~Boglione, R.~Boussarie, F.A.~Ceccopieri, I.O.~Cherednikov, 
  P.~Connor, M.G.~Echevarria, G.~Ferrera, J.~Grados Luyando, F.~Hautmann, H.~Jung, T.~Kasemets, K.~Kutak, J.P.~Lansberg, 
  A.~Lelek, G.~Lykasov, J.D.~Madrigal Martinez, P.J.~Mulders, E.R.~Nocera, E.~Petreska, C.~Pisano, R.~Placakyte, V.~Radescu, 
  M.~Radici, G.~Schnell, I.~Scimemi, A.~Signori, L.~Szymanowski, S.~Taheri Monfared, F.F.~van der Veken, H.J.~van Haevermaet, 
  P.~van Mechelen, A.A.~Vladimirov, S.~Wallon, Acta Phys. Polon. B {\bf 46}, 2501 (2015).
\bibitem{14} S.P.~Baranov, A.V.~Lipatov, N.P.~Zotov, Phys. Rev. D {\bf 78}, 014025 (2008). 
\bibitem{15} M.~Deak, F.~Schwennsen, JHEP {\bf 09}, 035 (2008). 
\bibitem{16} H.~Jung, S.P.~Baranov, M.~Deak, A.~Grebenyuk, F.~Hautmann, M.~Hentschinski, A.~Knutsson, 
  M.~Kraemer, K.~Kutak, A.V.~Lipatov, N.P.~Zotov, Eur. Phys. J. C {\bf 70}, 1237 (2010). 
\bibitem{17} PDG Collaboration, Chin. Phys. C {\bf 38}, 090001 (2014).
\bibitem{18} V.N.~Gribov and L.N.~Lipatov, Sov.J. Nucl. Phys. {\bf 15}, 438 (1972);\\
  L.N.~Lipatov, Sov. J. Nucl. Phys. {\bf 20}, 94 (1975);\\
  G.~Altarelli, G.~Parisi, Nucl. Phys. B {\bf 126}, 298 (1977);\\
  Yu.L.~Dokshitzer, Sov. Phys. JETP {\bf 46}, 641 (1977).
\bibitem{19} A.V.~Lipatov, M.A.~Malyshev, N.P.~Zotov, JHEP {\bf 05}, 104 (2012). % gamma + Q  
\bibitem{20} M.~Ciafaloni, Nucl. Phys. B {\bf 296}, 49 (1988);\\
  S.~Catani, F.~Fiorani, G.~Marchesini, Phys. Lett. B {\bf 234}, 339 (1990);\\
  S.~Catani, F.~Fiorani, G.~Marchesini, Nucl. Phys. B {\bf 336}, 18 (1990);\\
  G.~Marchesini, Nucl. Phys. B {\bf 445}, 49 (1995).
\bibitem{21} E.A.~Kuraev, L.N.~Lipatov, V.S.~Fadin, Sov. Phys. JETP {\bf 44}, 443 (1976);\\
  E.A.~Kuraev, L.N.~Lipatov, V.S.~Fadin, Sov. Phys. JETP {\bf 45}, 199 (1977);\\
  I.I.~Balitsky, L.N.~Lipatov, Sov. J. Nucl. Phys. {\bf 28}, 822 (1978).
\bibitem{22} F.~Hautmann, H.~Jung, Nucl. Phys. B {\bf 883}, 1 (2014).
\bibitem{23} J.~Kwiecinski, A.D.~Martin, P.~Sutton, Z. Phys. C {\bf 71}, 585 (1996).
\bibitem{24} B.~Andersson, G.~Gustafson, J.~Samuelsson, Nucl. Phys. B {\bf 467}, 443 (1996).  
\bibitem{25} M.~Hansson, H.~Jung, arXiv:hep-ph/0309009.
\bibitem{26} A.D.~Martin, W.J.~Stirling, R.S.~Thorne, G.~Watt, Eur. Phys. J. C {\bf 63}, 189 (2009).
\bibitem{27} C.~Peterson, D.~Schlatter, I.~Schmitt, P.~Zerwas, Phys. Rev. D {\bf 27}, 105 (1983).
\bibitem{28} P.~Bartalini, E.L.~Berger, B.~Blok, G.~Calucci, R.~Corke, M.~Diehl, Yu.~Dokshitzer, 
  L.~Fano, L.~Frankfurt, J.R.~Gaunt, S.~Gieseke, G.~Gustafson, D.~Kar, C.-H.~Kom, A.~Kulesza, E.~Maina, 
  Z.~Nagy, Ch.~Roehr, A.~Siodmok, M.~Schmelling, W.J.~Stirling, M.~Strikman, D.~Treleani, arXiv:1111.0469 [hep-ph].
\bibitem{29} H.~Abramowicz, P.~Bartalini, M.~Baehr, N.~Cartiglia, R.~Ciesielski, E.~Dobson, F.~Ferro, 
  K.~Goulianos, B.~Guiot, X.~Janssen, H.~Jung, Yu.~Karpenko, J.~Kaspar, J.~Katzy, F.~Krauss, P.~Laycock, 
  E.~Levin, M.~Mangano, Ch.~Mesropian, A.~Moraes, M.~Myska, D.~Moran, R.~Muresan, Z.~Nagy, T.~Pierog, 
  A.~Pilkington, M.~Poghosyan, T.~Rogers, S.~Sen, M.H.~Seymour, A.~Siodmok, M.~Strikman, P.~Skands, 
  D.~Treleani, D.~Volyanskyy, K.~Werner, P.~Wijeratne, arXiv:1306.5413 [hep-ph].
\bibitem{30} S.~Bansal, P.~Bartalini, B.~Blok, D.~Ciangottini, M.~Diehl, F.M.~Fionda, J.R.~Gaunt, 
  P.~Gunnellini, T.~Du Pree, T.~Kasemets, D.~Ostermeier, S.~Scopetta, A.~Siodmok, A.M.~Snigirev, A.~Szczurek, 
  D.~Treleani, W.J.~Waalewijn, arXiv:1410.6664 [hep-ph].
\bibitem{31} S.P.~Baranov, A.V.~Lipatov, N.P.~Zotov, Phys. Rev. D {\bf 89}, 094025 (2014). % DY
\bibitem{32} H.~Jung, M.~Kraemer, A.V.~Lipatov, N.P.~Zotov, Phys. Rev. D {\bf 85}, 034035 (2012). % bb LHC
\bibitem{33} H.~Jung, M.~Kraemer, A.V.~Lipatov, N.P.~Zotov, JHEP {\bf 01}, 085 (2011). % bb TeV
\bibitem{34} G.P.~Lepage, J. Comput. Phys. {\bf 27}, 192 (1978).
\bibitem{35} J.M.~Campbell, R.K.~Ellis, Phys. Rev. D {\bf 60}, 113006 (1999);\\
  J.M.~Campbell, R.K.~Ellis, C.~Williams, JHEP {\bf 1107}, 018 (2011);\\
  J.M.~Campbell, R.K.~Ellis, W.~Giele, Eur. Phys. J. C {\bf 75}, 246 (2015). 
\bibitem{36} S.~Dooling, F.~Hautmann, H.~Jung, Phys. Lett. B {\bf 736}, 293 (2014).
\bibitem{37} J.~Alwal, R.~Frederix, S.~Frixione, V.~Hirschi, F.~Maltoni, O.~Mattelaer, H.S.~Shao, 
  T.~Stelzer, P.~Torielli, M.~Zaro, JHEP {\bf 07}, 079 (2014).
  
\end{thebibliography}
\end{document}